\def\bold#1{\setbox0=\hbox{$#1$}%
     \kern-.025em\copy0\kern-\wd0
     \kern.05em\copy0\kern-\wd0
     \kern-.025em\raise.0433em\box0 }
\def\slash#1{\setbox0=\hbox{$#1$}#1\hskip-\wd0\dimen0=5pt\advance
       \dimen0 by-\ht0\advance\dimen0 by\dp0\lower0.5\dimen0\hbox
         to\wd0{\hss\sl/\/\hss}}
\documentstyle[12pt]{article}

\newlength{\dinwidth}
\newlength{\dinmargin}
\begin{document}

\def\lq{\left [}
\def\rq{\right ]}
\def\LL{{\cal L}}
\def\VV{{\cal V}}
\def\AA{{\cal A}}

\newcommand{\be}{\begin{equation}}
\newcommand{\ee}{\end{equation}}
\newcommand{\bea}{\begin{eqnarray}}
\newcommand{\eea}{\end{eqnarray}}
\newcommand{\nn}{\nonumber}
\newcommand{\dd}{\displaystyle}

\thispagestyle{empty}
\vspace*{1cm}
\rightline{December 1992}
\vspace{-42pt}
\rightline{BARI-TH/92-131}
\vspace*{4cm}
\begin{center}
  \begin{Large}
  \begin{bf}
           Study of the reactions  \\
          $ B \to D^* \pi \pi$ and $B\to D^* \rho \pi$\\
  \end{bf}
  \end{Large}
  \vspace{6mm}
  \begin{large}
Pietro Colangelo $^{a}$, Fulvia De Fazio $^{a}$, Giuseppe Nardulli $^{a,b}$\\
  \end{large}
\vspace*{1cm}
  \vspace{6mm}
$^{a}$ Istituto Nazionale di Fisica Nucleare, Sezione di Bari, Italy\\
  \vspace{4mm}
$^{b}$ Dipartimento di Fisica, Universit\'a
di Bari, Italy \\
\end{center}
\begin{quotation}
\vspace*{2cm}
\begin{center}
  \begin{bf}
  ABSTRACT
  \end{bf}
\end{center}
\vspace*{1cm}
We evaluate  the non leptonic B meson
decays $B \to D^* \pi \pi$ and
$B \to D^* \rho \pi$, in the factorization approximation and in the limit
of infinitely heavy quarks, assuming the dominance of intermediate positive
parity charmed resonances. We find that the branching ratios are of the order
$10^{-3}$.
\vspace{5mm}
\noindent

\end{quotation}
\vspace*{2cm}


\newpage
\setcounter{page}{1}

Positive parity charmed mesons have been recently investigated by a number of
authors \cite{IW,CNP,FL,KOR} in the framework of the Heavy Quark Effective
Theory (HQET). In particular in \cite{IW} it has been shown that the
semileptonic decays :
\begin{equation}
B \to D^{**} \ell \nu
\end{equation}
can be described in terms of two universal form factors,
$\tau_{1/2}(\omega)$ and
$\tau_{3/2}(\omega)$, where $\omega=v\cdot v'$ and $ v^{\mu}$, $v'^{\mu}$ are
the B and $D^{**}$ meson four-velocities respectively. In eq.(1) $D^{**}$
 is one of the four positive parity charmed mesons:
 $2^+_{3/2}$, $1^+_{3/2}$, $1^+_{1/2}$,
$0^+_{1/2}$, that one expects in the infinite heavy quark mass limit; here we
employ the notation $J^P_{s_{\ell}}$, where $s_{\ell}=1/2$ or $3/2$
is the total angular momentum of the light degrees of freedom. The form factor
$\tau_{3/2}(\omega)$ describes the decays into the $2^+_{3/2}$ and $1^+_{3/2}$
states, whereas $\tau_{1/2}(\omega)$ is related to the $1^+_{1/2}(\omega)$
and $0^+_{1/2}$
states. Both form factors have been estimated in \cite{CNP} by using the QCD
sum rules approach.

The $2^+_{3/2}$ state has been observed experimentally \cite{PDG} with a mass
of $2460 \hskip 3pt MeV$: it is denoted by $D_2^*(2460)$.
The $1^+_{3/2}$ meson has to be
basically identified with $D_1(2420)$\footnote{It is likely however that $D_1(
2420)$ also contains an admixture of $1^+_{1/2}$; for discussions see
\cite{FL} and \cite{KOR}.}.
As shown in refs.\cite{FL,KOR}  they are both narrow
$(\Gamma \leq 20 \hskip 3pt MeV)$
since their strong decays occur in $D$-wave, in contrast with the states
$D_0$ (the $0^+_{1/2}$ state) and $D'_1$ (the $1^+_{1/2}$ state)
that can also decay by
$S$-wave.
On the experimental side, some evidence has been gathered on the semileptonic
decays (1) \cite{CLEO,ARGUS}, but it is not yet conclusive. A different way
to study
the transition $B \to D^{**}$ is by the non leptonic reactions \cite{Stone} :
\begin{eqnarray}
B &\to& D^* \pi \pi \\
B &\to& D^* \rho \pi
\end{eqnarray}
which can occur by the intermediate production of positive parity resonances
with
subsequent decay into $D^* \pi$.
Examples of these processes  are as follows:
\begin{eqnarray}
B^- &\to& D^{**0} \pi^- \to D^{*+} \pi^- \pi^- \\
B^- &\to& D^{**0} \rho^- \to D^{*+} \pi^- \rho^- \\
{\bar B^0} &\to& D^{*0} \pi^+ \pi^-
\end{eqnarray}
where in the last case we could have either $D^{*0} \rho^0$ or $D^{**+} \pi^-$
as intermediate resonant states.

In this letter we wish to study the processes (4-6) in the framework of the
factorization approximation. This approach has first been proposed
by Feynman \cite{FEY}, and then  has been extensively applied by Bauer, Stech
and Wirbel (BWS) \cite{BWS} and by a number of other authors
\footnote{We do not include final state strong
interaction effects due to the lack of data on the phase shifts. For an
analysis of such effects in $B \to D^* \pi$ and $B \to D D$ decays see
ref.~\cite{Tan}. For other studies of non leptonic $B$ meson decays see
refs.~\cite{BWS,CHAU}
and references therein.}. As well known, in this approximation
one considers matrix elements of the weak non leptonic effective hamiltonian
\begin{equation}
H_{NL}= {G \over \sqrt 2} V^*_{cb} V_{ud}
:a_1 \hskip 3pt {\bar c} \gamma_\mu (1-\gamma_5) b \hskip 2pt
{\bar d} \gamma^\mu (1-\gamma_5) u \hskip 4pt + \hskip 4pt
a_2 \hskip 3pt {\bar d} \gamma_\mu (1-\gamma_5) b \hskip 2pt
{\bar c} \gamma^\mu (1-\gamma_5) u :
\end{equation}
and evaluates them by inserting the hadronic vacuum state between the $V-A$
quark currents appearing in (7); the constants $a_1$ and $a_2$ are given by:
\begin{eqnarray} \label{eq8}
a_1 &=& c_1 + {c_2 \over N_c} \\
a_2 &=& c_2 + {c_1 \over N_c}
\end{eqnarray}
where $c_1=1.1$ and $c_2=-0.24$ are Wilson coefficients evaluated at the
$b-$quark mass scale. As for $N_c$, experimental data for $B \to D \pi$ and
 $B \to D \rho$ seem to favour the value of the BWS model~\cite{BWS}
$N = \infty$ (see e.g.~\cite{Tan}) instead of the value $N_c=3$; we shall
comment on this point later on.

A test of factorization has been worked out in ref. \cite{SB}, where
the experimental ratio between the
${\bar B^0} \to D^{*+} \pi^-$ non leptonic width and semileptonic spectrum
${d \Gamma ({\bar B^0} \to D^{*+} \ell^- {\bar \nu}) \over d q^2}$,
evaluated at
$q^2=m_\pi^2$, has been used. If factorization
works well, this quantity is a model independent prediction;
the result shows that,
 within the accuracy of the present
experimental data, factorization is correct at the $25 \%$ level.

In order to compute the hadronic matrix elements of $H_{NL}$ we need the weak
current matrix elements between hadron states. In the infinite heavy quark mass
limit they can be written as follows~\cite{IW,IW1,Falk1}:

\begin{eqnarray} \label{eq9}
<D(v')|V_\mu(0)|B(v)> &=& {\sqrt {m_B m_D}} \hskip 4pt \xi(v\cdot v')
 \hskip 4pt (v+v')_\mu \\
<D^*(v',\epsilon)|(V-A)_\mu(0)|B(v)> &=& {\sqrt {m_B m_D}}  \hskip 4pt
\xi(v\cdot v')  \hskip 4pt
[\hskip 4pt i \epsilon_{\mu \nu \alpha \beta} \epsilon^{*\nu} v^\alpha v'^\beta
 \nonumber\\
&-& \epsilon^*_\mu (1+v\cdot v') + (\epsilon^*\cdot v) \hskip 4pt v'_\mu] \\
<D_0(v')|A_\mu(0)|B(v)> &=& - 2 {\sqrt {m_B m_{D^{**}}}}  \hskip 4pt
\tau_{1/2} (v\cdot v')  \hskip 4pt (v-v')_\mu \\
<D'_1(v',\epsilon)|(V-A)_\mu(0)|B(v)> &=& {\sqrt {m_B m_{D^{**}}}}
 \hskip 4pt \tau_{1/2}(v\cdot v') \hskip 4pt
[2 \hskip 4pt (v \cdot v' -1) \hskip 4pt \epsilon^*_\mu  \nonumber \\
&+& 2 \hskip 4pt (\epsilon^*\cdot v) v'_\mu -
i \epsilon_{\mu\nu\alpha\beta} \hskip 4pt \epsilon^{*\nu}
(v+v')^\alpha (v-v')^\beta] \\
<D_1(v',\epsilon)|(V-A)_\mu(0)|B(v)> &=& {\sqrt {m_B m_{D^{**}}}}
 \hskip 4pt \tau_{3/2}(v\cdot v') \hskip 4pt
\Big \{ [{ (1 - (v \cdot v')^2 ) \over \sqrt 2} \hskip 4pt
\epsilon^*_\mu  \nonumber \\
&+& { (\epsilon^*\cdot v) \over \sqrt 2} [- 3 \hskip 4pt v_\mu +
 (v \cdot v' -2) \hskip 4pt v'_\mu ]  \nonumber \\
&+&
i \hskip 4pt {(v \cdot v' - 1) \over 2 \sqrt 2} \epsilon_{\mu\nu\alpha\beta}
\epsilon^{*\nu} (v+v')^\alpha (v-v')^\beta] \Big \} \\
<D^*_2(v',\epsilon)|(V-A)_\mu(0)|B(v)> &=& {\sqrt {m_B m_{D^{**}}}}
 \hskip 3pt \tau_{3/2}(v\cdot v') \hskip 3pt
[ \hskip 3pt i \hskip 4pt {{\sqrt 3} \over 2} \epsilon_{\mu\alpha\beta\gamma}
\epsilon^{*\alpha\nu} v_\nu (v+v')^\beta (v-v')^\gamma \nonumber \\
&-& {\sqrt 3} (v \cdot v' +1) \epsilon^*_{\mu\alpha} v^\alpha \hskip 3pt +
\hskip 3pt {\sqrt 3} \epsilon^*_{\alpha\beta} v^\alpha v^\beta v'_\mu ]
\end{eqnarray}
where $V^\mu= {\bar c}_{v'} \gamma^\mu b_v$,
$A^\mu= {\bar c}_{v'} \gamma^\mu \gamma_5 b_v$,
$c_{v'}$ and  $b_v$ are quark operators in HQET,  $m_{D^{**}}$
is the mass of the
positive parity charmed resonances, and $\xi(v \cdot v')$,
$\tau_{1/2}(v \cdot v')$ and
$\tau_{3/2}(v \cdot v')$ are the Isgur-Wise universal form factors~\cite{IW}.
For the form factors $\tau_{1/2}(v \cdot v')$ and $\tau_{3/2}(v \cdot v')$ we
take the results of Ref. \cite{CNP},
where these form factors have been obtained by 3-point function QCD sum rules
in the $m_Q \to \infty$ limit. The results of this approach are in reasonable
agreement with an estimate based on the non relativistic potential model
\cite{IW};
 in particular at $v \cdot v'=1.3$ which is
the relevant value for $v \cdot v'={m^2_B + m^2_{D^{**}} -q^2 \over2m_B
m_{D^{**}}}|_{q^2 \simeq m^2_{\pi}, m^2_{\rho}}$ in our case, we use
$\tau_{1/2}(v \cdot v')
=0.20$ and $ \tau_{3/2}(v \cdot v')=0.19$.

We have also to consider the matrix elements of the currents between the vacuum
and the charmed resonances in the $m_Q \to \infty$ limit:
\begin{eqnarray}
<0| V^{\mu}(0)| D_0(p)> &=& i f^{(+)} /\sqrt{m_c} p^{\mu}  \\
<0| A^{\mu}(0)| D'_1(\epsilon, p)> &= &f^{(+)} \sqrt{m_c}  \hskip 3pt
\epsilon^{\mu}
\end{eqnarray}
where $f^{(+)}$ depends only logarithmically on the heavy quark mass and has
been determined in ref.~\cite{CNP} by QCD sum rules ($m_c=1.35 \hskip 3pt
GeV$):
\begin{equation}
f^{(+)}\simeq \hskip 3pt 0.46 \hskip 3pt GeV^{3/2}.
\end{equation}
In contrast, the matrix elements between the vacuum and the $s_{\ell}=3/2$
resonances vanish in the $m_Q \to \infty$ limit \cite{CNP}.

Finally, we have to consider the weak current matrix elements between $B$ and a
light meson $\pi, \rho$; they can be written as follows:
\begin{eqnarray}
<\pi(p')| V^\mu| B(p)> &=& (p^\mu +p'^\mu - {(m^2_B -m^2_\pi) \over q^2} q^\mu)
F_1(q^2) \nonumber \\
& + & {(m^2_B -m^2_\pi) \over q^2} q^\mu F_0(q^2) \\
<\rho(p',\epsilon^*)| V^\mu - A^\mu |B(p)> &=& \epsilon^{\mu\nu\rho\sigma}
\epsilon^*_\nu p_\rho p'_\sigma { 2 V(q^2) \over (m_B + m_\rho)} \nonumber \\
& +& i \Big\{ \epsilon^{*\mu} (m_B + m_\rho) A_1(q^2) +
(\epsilon^*\cdot p) { A_2(q^2)\over (m_B +m_\rho)} (p+p')^\mu \nonumber \\
&+& (\epsilon^*\cdot p) { q^\mu \over q^2} [(m_B +m_\rho) A_1(q^2) \nonumber \\
&-& (m_B - m_\rho) A_2(q^2) -2m_\rho A_0(q^2)] \Big\}
\end{eqnarray}
where
$F_0(0)=F_1(0)$ and $2m_{\rho} A_0(0)= (m_B +m_\rho) A_1(0) -
(m_B -m_{\rho}) A_2(0)$.

For the $q^2$ dependence of the various form factors in (18) and (19) we have
assumed a simple pole formula, i.e. $F(q^2)=F(0)/(1-q^2/m^2)$ with $ m=5.32
\hskip 3pt GeV$ (for $F_1$ and $V$), $m=5.99 \hskip 3pt GeV$
(for $F_0$), $m=5.73 \hskip 3pt GeV$ (for $A_1$ and $A_2$). As for their
values at $q^2=0$ we take the results of an analysis of the semileptonic $B$
decays into light mesons, obtained assuming both chiral symmetry and HQET
\cite{CG}: $F_1(0)=0.53, \hskip 3pt A_1(0)=0.21,  \hskip 3pt  A_2(0)=0.20,
 \hskip 3pt V(0)=0.62$.
These results are obtained by an effective lagrangian which incorporates the
symmetries for light and heavy quarks, generalizing the results of ref.
\cite{W} to all the semileptonic B decays into light particles. The quoted
values are obtained without $1/m_Q$ corrections, to be consistent with the
choices (15)-(17); the role of possible $1/m_Q$ terms will be discussed later
on.

The computed widths for the various $B^- \to D^{**0} \pi^-, \rho^-$ decay
channels are collected in Table I and Table II. It is worth observing that
the $a_2$ contributions, when present, are quite sizeable,
mainly due to the small numerical value of the universal form factor
$\tau_{1/2}$.
On the other hand, the $a_2$ contributions
are absent in the case of $B \to D^{**}(J_{3/2}^+) \pi, \rho$ since, as
observed above, the weak current matrix elements between the vacuum and the
$3/2$ doublet vanish in the limit of infinitely heavy quarks. This allows
us to give the following predictions for the ratios
${\Gamma(B^- \to D^{**0}(2^+_{3/2})  \hskip 3pt \pi^-) \over
\Gamma(B^- \to D^{**0}(1^+_{3/2})  \hskip 3pt \pi^-)}$ and
${\Gamma(B^- \to D^{**0}(2^+_{3/2})  \hskip 3pt \rho^-) \over
\Gamma(B^- \to D^{**0}(1^+_{3/2}) \hskip 3pt \rho^-)}$:
\begin{equation}
{\Gamma(B^- \to D^{**0}(2^+_{3/2})  \hskip 3pt \pi^-) \over
\Gamma(B^- \to D^{**0}(
1^+_{3/2})  \hskip 3pt \pi^-)}={4z  \hskip 3pt (1+\sqrt{z})^2 \over
(1- \sqrt{z})^2 [ (1+ \sqrt{z})^2-y]^2}
\end{equation}
and
\begin{equation}
{\Gamma(B^- \to D^{**0}(2^+_{3/2})  \hskip 3pt \rho^-) \over
\Gamma(B^- \to D^{**0}(1^+_{3/2}) \hskip 3pt \rho^-)}
= {4[y (3\omega +2) +(\omega-1)(1+ \sqrt{z})^2] \over [y(
3\omega-7)+ 5z\omega- 2\sqrt{z}\omega^2 +5\omega- 6\sqrt{z}\omega +3z]}
\end{equation}
( $z=\big({m_{D^{**}} \over m_B}\big)^2$,
$y=\big({m_\pi, m_\rho \over m_B}\big)^2$
and $\omega={1+z-y\over 2 \sqrt{z}}$)
which are independent of $\tau_{3/2}$;
however, the experimental test of these relations is a difficult task.

In order to evaluate the transitions (4-6), we use the narrow width
approximation for the positive parity charmed resonances, which is a good
assumption for $D^*_2(2460)$ and $D_1(2420)$. As for $D'_1$ ($D_0$ does not
decay into $D^* \pi$), an estimate based on a costituent quark model \cite{IS}
and HQET in the chiral limit \cite{FL} gives $\Gamma(D'_1)\simeq \hskip 3pt
80  \hskip 3pt MeV$, a value compatible with the narrow width approximation.
We also observe that, in the limit $m_{D_1}=m_{D'_1}$ and
$\Gamma_{D_1}=\Gamma_{D'_1}\simeq 0$ the mixing angle between
$D_1$ and $D'_1$ does not affect the final results.

The branching ratios for the $D^{**}$ decays into $D^* \hskip 3pt \pi$
have been
evaluated in ref.~\cite{FL} in the framework of the Heavy Quark Effective
Chiral Perturbation Theory, with the results:
\begin{eqnarray}
BR \hskip 3pt  (D'^0_1 \to D^{*+} \hskip 3pt \pi^-) &=& 2/3 \\
BR  \hskip 3pt (D^0_1(2420) \to D^{*+} \hskip 3pt \pi^-) &=& 2/3 \\
BR  \hskip 3pt (D^{*0}_2(2460) \to D^{*+} \hskip 3pt \pi^-) &=& 0.20 \hskip
6pt.
\end{eqnarray}

Using these branching ratios and the results in Table I and Table II
(with $a_1=1.1$ and
$a_2=-0.24$) we estimate for the transitions (4-6) the following BR's
($\tau_B=1.2 \hskip 3pt ps$):
\begin{eqnarray}
BR \hskip 3pt (B^- \to D^{*+} \hskip 3pt \pi^- \hskip 3pt \pi^-) &=& 1
\times 10^{-3}\\
BR  \hskip 3pt ({\bar B^0} \to D^{*0} \hskip 3pt \pi^+ \hskip 3pt \pi^-) &=& 6
\times 10^{-4}\\
BR  \hskip 3pt (B^- \to D^{*+} \hskip 3pt \pi^- \hskip 3pt \pi^-
\hskip 3pt \pi^0 ) &=& 2  \times 10^{-3} \hskip 6pt .
\end{eqnarray}
In computing Eq.(27) the channel
${\bar B^0} \to D^{*0} \hskip 3pt \rho^0$, which is a pure $a_2$ process,
 has been included incoherently.

Let us conclude with some comments on the accuracy of the results
Eqs.(26-28). First, it should be
observed that, in considering the processes (4-6), we have not taken into
account  non resonant pion production, since we presume dominance of resonant
behaviour.
Second, we have obtained the results (26-28) in the limit where both the beauty
and the charmed quark masses are taken to infinity.
The accuracy of the predictions obtained in this limit cannot be assessed by
general arguments, since we know that in some cases the $1/m_Q$ corrections
may be relevant (for instance in the calculation of the B leptonic decay
constant $f_B$ \cite{AB}) and in other cases they are small \cite{LUKE}.
Moreover we cannot present a complete analysis of such corrections for the $B
\to D^{**}$ semileptonic transitions, because a study of these processes at all
orders in $1/m_Q$ has been performed only for the states $D_0$ and $D'_1$
\cite{OVC}. We can nevertheless estimate the size of the possible errors by
looking at one particular channel, i.e. $B^- \to D^{**0}(0^+_{1/2}) \pi^-$. In
this case we can compare the entry in Table I with the result obtained
including $1/m_Q$ corrections. The last one follows from the matrix elements :
$<D^{**0}|A_{\mu}|B^->$ and $<0|V_{\mu}|D^{**0}>$ computed at all the orders in
$1/m_Q$ by QCD sum rules \cite{OVC}, together with the matrix element: $<\pi|
V^{\mu}|B>$ computed in ref. \cite{CG} by including $1/m_Q$ corrections (in
this case one should use in (19) $F_1(0)=0.89$ instead of $0.53$ ).
We obtain for
B.R.($B^- \to D^{**0}(0^+_{1/2}) \pi^-$) a result which differs by less
than 10\%
from the asymptotic ($m_Q \to \infty$) value reported in Table I, which
provides a perhaps optimistic estimate of the theoretical errors introduced by
the heavy quark mass limit. If we observe that uncertainties in the
phenomenological value of $N_c$ (see eqs.(8,9)) would strongly affect
only the $B^-$ decays into the  $0^+_{1/2}$ state, which however does not
decay into $D^* \pi$ in the infinite heavy quark limit, we can conclude that
our predictions (26)-(28) should provide reasonable estimates of the non
leptonic $B^-$ decays into charmed multipion final states.

\vspace*{4cm}
{\bf Acknowledgments }

We thank S.Stone for having suggested this calculation, and N.Paver for useful
discussions.

\newpage
\begin{center}
  \begin{Large}
  \begin{bf}
  Table Captions
  \end{bf}
  \end{Large}
\end{center}
  \vspace{5mm}
\begin{description}
\item [Tab. I]  Widths and branching ratios (B.R.)
of $B^- \to D^{**0}(J^P_{s_\ell}) \hskip 3pt
\pi^-$. The $B^-$ lifetime is $\tau_B=1.2 \times 10^{-12}$ sec.
The branching ratios are obtained for $a_1=1.1$, $a_2=-0.24$ and $V_{cb}=0.
045$.
\vspace{5mm}
\item [Tab. II] Widths and branching ratios (B.R.) of
$B^- \to  D^{**0}(J^P_{s_\ell}) \hskip 3pt
\rho^-$. The $B^-$ lifetime is $\tau_B=1.2 \times 10^{-12}$ sec.
The branching ratios are obtained for $a_1=1.1$, $a_2=-0.24$ and $V_{cb}=0.
045$.
\end{description}
\newpage
\begin{table}
\begin{center}
\begin{tabular}{l c c }
& {\bf Table I} & \\ & & \\
 \hline \hline
$J^P_{s_\ell}$ & Width (GeV)  & B.R. \\ \hline
$2^+_{3/2}$ & $2.64  \hskip 3pt a_1^2 \hskip 3pt 10^{-16} \hskip 3pt
\big( {V_{cb} \over 0.045}\big)^2$ & $ 6 \hskip 3pt 10^{-4}$\\ \hline
$1^+_{3/2}$ & $2.02  \hskip 3pt a_1^2 \hskip 3pt 10^{-16} \hskip 3pt
\big( {V_{cb} \over 0.045}\big)^2$ & $ 4 \hskip 3pt 10^{-4}$\\ \hline
$1^+_{1/2}$ & $1.01  \hskip 3pt (a_1^2 + 48.1 a_2^2) \hskip 3pt 10^{-16} \hskip
3pt \big( {V_{cb} \over 0.045}\big)^2$ & $ 6 \hskip 3pt 10^{-4}$\\ \hline
$0^+_{1/2}$ & $0.87  \hskip 3pt (a_1 - 14.7 a_2)^2 \hskip 3pt 10^{-16} \hskip
3pt \big( {V_{cb} \over 0.045}\big)^2$ & $ 3 \hskip 3pt 10^{-3}$\\ \hline
\hline
\end{tabular}
\end{center}
\end{table}
\vspace{1cm}
\begin{table}
\begin{center}
\begin{tabular}{l c c }
& {\bf Table II} & \\ & & \\
 \hline \hline
$J^P_{s_\ell}$ & Width (GeV)  & B.R. \\ \hline
$2^+_{3/2}$ & $5.08 \hskip 3pt a_1^2 \hskip 3pt 10^{-16}
\hskip 3pt \big( {V_{cb} \over 0.045}\big)^2$ & $ 1 \hskip 3pt 10^{-3}$ \\
\hline
$1^+_{3/2}$ & $4.8 \hskip 3pt a_1^2 \hskip 3pt 10^{-16}
\hskip 3pt \big( {V_{cb} \over 0.045}\big)^2$ & $ 1 \hskip 3pt
10^{-3}$\\ \hline
$1^+_{1/2}$ & $2.5 \hskip 3pt (a_1^2 + 17.9 a_2^2) \hskip 3pt 10^{-16}
\hskip 3pt \big( {V_{cb} \over 0.045}\big)^2$ &
$ 1 \hskip 3pt 10^{-3}$\\ \hline
$0^+_{1/2}$ & $1.96 \hskip 3pt (a_1 + 3.2 a_2)^2 \hskip 3pt 10^{-16}\hskip 3pt
\big( {V_{cb} \over 0.045} \big)^2$ & $ 4 \hskip 3pt 10^{-5}$\\ \hline
\hline
\end{tabular}
\end{center}
\end{table}
\vspace{1cm}
\end{document}